\documentclass[fleqn,usenatbib]{rasti}
\usepackage{hyperref}
\usepackage{subcaption}
\usepackage{booktabs}

\newcommand{\oiii}{[O\,\textsc{iii}]}
\usepackage{newtxtext,newtxmath}
\usepackage{fontawesome}

\usepackage[T1]{fontenc}

\DeclareRobustCommand{\VAN}[3]{#2}
\let\VANthebibliography\thebibliography
\def\thebibliography{\DeclareRobustCommand{\VAN}[3]{##3}\VANthebibliography}

\usepackage{graphicx}	
\usepackage{amsmath}	

\title[Simultaneous PSF Matching]{Simultaneous PSF Matching via Dual Kernel Optimization}

\author[Carter Lee Rhea et al.]
{Carter Lee Rhea,$^{1,2}$\thanks{E-mail: carter.rhea@dragonfly1000.com}
Pieter van Dokkum,$^{1,3}$ Roberto Abraham,$^{1,4,5}$ Imad Pasha,$^{1,3}$ \newauthor Steven R. Janssens,$^{1}$ William P. Bowman,$^{1,3}$ Deborah Lokhorst,$^{1,6}$ Seery Chen$^{1,4,5}$, Qing Liu$^{7}$
\\
$^{1}$Dragonfly Focused Research Organization, 150 Washington Avenue, Santa Fe, NM 87501, USA\\
$^{2}$Centre de recherche en astrophysique du Québec (CRAQ), Québec, QC G1V 0A6, Canada\\
$^{3}$Astronomy Department, Yale University, 219 Prospect St, New
Haven, CT 06511, USA\\
$^{4}$David A. Dunlap Department of Astronomy \& Astrophysics,University of Toronto, 50 St. George Street, Toronto, ON M5S3H4,Canada\\
$^{5}$Dunlap Institute for Astronomy \& Astrophysics, University of Toronto, 50 St. George Street, Toronto, ON M5S3H4, Canada\\
$^{6}$NRC Herzberg Astronomy \& Astrophysics Research Centre, 5071 West Saanich Road, Victoria, BC V9E 2E7, Canada\\
$^{7}$Leiden Observatory, Leiden University, P.O. Box 9513, 2300 RA Leiden, The Netherlands\\
}

\date{Accepted XXX. Received YYY; in original form ZZZ}

\pubyear{\the\year{}}

\begin{document}
\label{firstpage}
\pagerange{\pageref{firstpage}--\pageref{lastpage}}
\maketitle

\begin{abstract}
Standard point spread function (PSF) matching algorithms consider one image as the reference and aim to find a convolution kernel such that, when applied to the reference image, the PSFs of the reference and non-reference images are matched. This method works well if one image is a degraded version of the other.  However, this is not always the case: if the PSFs of the two images have different orientations or shapes, a simple convolution of one of the images does not lead to a satisfactory match. While this can be trivially solved by convolving the first PSF with the second and vice-versa, this results in unnecessarily large PSFs.
In this paper, we present a new algorithm, Dual Kernel Optimization (DKO), that simultaneously solves for two convolution kernels, one for each image, such that the PSFs match and the sizes of the kernels are minimized. This results in a minimally-degraded final set of PSFs. Since this problem is highly degenerate, we use Stochastic Gradient Langevin Dynamics (SGLD) which rigorously explores the parameter space and converges on the globally-optimal solution.
We discuss the algorithmic challenges for this problem and the modifications to standard SGLD used to constrain it. 
Finally, we include a discussion on the application of this methodology to real-world images, including cases where the PSF varies over the field of view. 
\end{abstract}




\section{Introduction}
The point spread function (PSF) plays a crucial role in all astronomical observations, as it describes how incident light is spread out on a detector (e.g., \citealt{racine_telescopic_1996}). In an ideal system, the PSF would be a delta function; however, in practical optical systems, the PSF  is a complex function of the optical system and, for ground-based telescopes, the aberrations induced by the atmosphere (e.g., \citealt{anderson_psfs_2006}; \citealt{liaudat_point_2023}; \citealt{berge_point_2012}). Since the PSF can be affected by transient effects, such as the focus and atmospheric conditions, it can change from exposure to exposure; moreover, PSFs can vary as a function of the position on the detector. Therefore, in order to properly treat multiple images, whether for stacking purposes (e.g., \citealt{alard_method_1998}; \citealt{alard_image_2000}), detecting transient events (e.g., \citealt{sako_sloan_2007}), or accurate photometric measurements (e.g., \citealt{turri_optimal_2017}), it is important to homogenize the PSFs. Moreover, it is preferable to keep the final homogenized PSF as compact as possible so as to not lose spatial information. Therefore, a considerable amount of scientific effort has focused on matching PSFs from one image to another  (see, e.g., \citealt{mancone_pygfit_2013}; \citealt{merlin_t-phot_2015}; \citealt{aniano_common-resolution_2011}; \citealt{price_pan-starrs_2019}).

The majority of methods for matching the PSFs of two images implicitly or explicitly assume that one image is the degraded version of the other, so they solve the following equation:
\begin{equation}\label{eqn:simple-conv}
    \mathrm{PSF}_1 \otimes K = \mathrm{PSF}_2 ,
\end{equation}
where $K$ is the convolutional kernel that will match PSF$_1$ and PSF$_2$ and $\otimes$ is the convolution operator.
The most common method for solving this equation, frequently referred to as the Alard Method (\citealt{alard_image_2000}), involves rewriting $K$ as a series of basis functions, $K=\sum_{i}\alpha_i K_i$, where the $K_i$ functions are selected from a set of analytic basis functions such as delta functions or Gaussians. Using standard optimization algorithms, such as the method of least squares, $K$ can often be recovered to a suitable level of reconstruction error. Some authors have explored regularization methods to stabilize the solutions in the case of degenerate solutions (e.g., \citealt{becker_regularization_2012}). Other authors have avoided selecting basis functions by applying a least squares analysis on the pixel  values of the convolutional kernel directly (e.g., \citealt{bramich_new_2008}). While all of these methods have shown to be effective, they assume that Eq.\ \ref{eqn:simple-conv} has a solution by implicitly assuming that $\mathrm{PSF}_2$ is a degraded version of $\mathrm{PSF}_1$.

Other authors have explored methods to solve for both PSF$_1$ and PSF$_2$ without assuming any degeneracy between the two such as: the LSST-style delta function (\citealt{becker_regularization_2012}; \citealt{bramich_new_2008}; \citealt{alard_method_1998}; \citealt{alard_image_2000}), or cross-correlation (\citealt{yuan_astronomical_2008}). These approaches have been shown to generally effective at resulting in similar convolved PSFs; however, they can leave systematic residuals.

Therefore, in this paper, we present Dual Kernel Optimization (DKO), a framework for simultaneous PSF matching that does not assume either image is a degraded version of the other. Instead, we derive two convolutional kernels simultaneously, which produce matched PSFs after convolution of both images with their respective kernels while also forcing the kernels to be as compact as possible. The method was developed for continuum subtraction of narrow band imaging with the MOTHRA telescope, but is generally applicable.
In \S\,\ref{sec:methods}, we outline the mathematical description of the problem and our proposed methodology, and show how the algorithm can be applied to toy problems. In \S\,\ref{sec:application} we demonstrate how the methodology can be used in real world scenarios. Finally, we discuss additional applications and summarize the results in \S\,\ref{sec:conclusions}. All relevant code and documentation can be found at \href{https://github.com/DragonflyTelescope/dfpsf}{\faicon{github}\texttt{DragonflyTelescope/dfpsf}}.

\section{Methods}\label{sec:methods}
In this section we first describe the mathematics of our methodology. Then we apply the method to several toy problems.

\subsection{Algorithmic Description}
Rather than searching for a single convolutional kernel that will transform a source PSF into a target PSF, DKO seeks two convolutional kernels, $K_1$ and $K_2$, that satisfy
\begin{equation}\label{eqn:primary}
    \mathrm{PSF}_1 \otimes K_1 = \mathrm{PSF}_2 \otimes K_2.
\end{equation}
We note that this equation is highly degenerate with the trivial solution being $K_1=\mathrm{PSF}_2$ and $K_2=\mathrm{PSF}_1$, and we use this fact in our methodology: we test our initial solutions to this. While the trivial solution works, it results in PSFs with a large FWHM in the convolved images, which are generally not desirable since the spatial information is degraded. Therefore, we derive a method which solves for the smallest possible kernels that satisfy Eq.\ \ref{eqn:primary}.

\subsubsection{Parameterization of the Kernels}

We parametrize the convolutional kernel as a mixture of  Gaussian basis functions:
\begin{equation}
    K_\mathrm{total}(x,y) = \sum_{i=1}^{N}w_i K_i(x,y) ,
\end{equation}
where $N$ is the number of components and $w$ is a fitted weight parameter. In the  case that we are only solving a single base function, we set $w_1=1$.
We note that all free parameters are fit simultaneously for each individual Gaussian component. 

Each basis function is given by a 2D elliptical Gaussian:
\begin{equation}\label{two2gaussian}
K(x, y) = \exp\left[ -\frac{1}{2} \left( a(x - x_0)^2 + b(x - x_0)(y - y_0) + c(y - y_0)^2 \right) \right],
\end{equation}
where 
\begin{equation*}
    a = \frac{\cos{}^2(\theta)}{\sigma_x^2} + \frac{\sin{}^2(\theta)}{\sigma_y^2} ,
\end{equation*}
\begin{equation*}
    b = 2\Bigg(\frac{1}{\sigma_x^2} - \frac{1}{\sigma_y^2}\Bigg) \sin{}(\theta) \cos{}(\theta) ,
\end{equation*}
\begin{equation*}
    c = \frac{\sin{}^2(\theta)}{\sigma_x^2} + \frac{\cos{}^2(\theta)}{\sigma_y^2}.
\end{equation*}
We define the following additional values: 
$\sigma_x = \frac{\sigma}{\sqrt{q}}$, $\sigma_y = \sigma \sqrt{q}$, $x = \mu + dx$, $y = \mu + dy$.
In this parametrization $\mu$ is set to the center of the kernel leaving the following free variables: $dx, dy, \sigma, q,$ and $\theta$. 
$dx$ and $dy$ represent the displacement in the x and y positions, respectively. $\sigma$ is the width of the Gaussian, $q$ is the ellipticity of the Gaussian, and $\theta$ is the counter-clockwise position angle in radians.
In order to reduce the number of components, we set $dx$ and $dy$ of $K_2$ equal to zero; despite this, the convolution still captures small positional shifts in the PSFs that could be due to small astrometric errors since we leave $dx$ and $dy$ of $K_1$ as free parameters.

\subsubsection{Fitting Framework}

In order to solve Equation \ref{eqn:primary}, we use Stochastic Gradient Langevin Dynamics (SGLD; e.g., \cite{welling_bayesian_2011}; \cite{teh_consistency_2016}; \citealt{brosse_promises_2018}). This method, rooted in Langevin dynamics (e.g., \citealt{turq_brownian_1977}), aims to solve the following equation:
\begin{equation}\label{eqn:langevin}
    \phi_i^{t+\tau} = \phi_i^t + \tau \nabla_i \pi(x) + \sqrt{2\tau T}\zeta,
\end{equation}
where $\phi$ is the solution, $i$ indexes the parameter of interest, $t$ is the solution time step, $\tau$ is the step size, $T$ is the Langevin temperature, $\pi(x)$ is the loss function, and $\zeta$ is a noise value sampled from $\mathcal{N}(0,1)$. By adopting this formulation, we allow for some exploration of the parameter space while exploiting the gradient of the loss function to drive the parameters towards their local minima. Since this is a gradient-based method, we are required to have a differentiable loss function that can be written analytically.

The reformulation in Equation (2) introduces a highly degenerate optimization problem. DKO resolves this degeneracy through a loss function that simultaneously enforces PSF agreement, compact kernels, and smooth solutions, while SGLD provides robust exploration of the resulting parameter space. 
All  penalty terms are dynamically scaled as a function of the $\chi^2$ term.
Therefore, we can write the total loss function as:
\begin{equation}\label{eqn:loss}
    \pi(\vec{x}) =  \chi^2(\vec{x}) ( 1 + \beta_1 \xi_{\rm size}(\vec{x}) + \beta_2 \xi_{\rm smooth}(\vec{x}) + \beta_3 \xi_{L^2}(\vec{x})),
\end{equation}
where $\vec{x}$ represents our free parameters.
By dynamically scaling the penalty terms as a function of the $\chi^2$ loss term, we ensure that this term, also called the image match loss, is prioritized over the penalty terms until it has converged. The $\beta$ terms determine the relative importance of the penalty terms compared to the image match loss. We find that $\beta_1 = 0.1$, $\beta_2 = 0.2$, and $\beta_3 = 0.05$ yield a good trade-off between the different penalty terms.

The weighted $\chi^2$ term is as follows:
\begin{equation}\label{eqn:chi}
    \chi^2 = \sum \Big(\mathrm{PSF}_1\otimes K_1 - \mathrm{PSF}_2 \otimes K_2 \Big)^2\Bigg( \frac{1}{n_{\mathrm{pixels}}} \Bigg),
\end{equation}
where  $n_{\text{pixels}}$ is the number of pixels in the PSF image.

The regularization is composed of three terms: a size penalty term, a smoothness penalty term, and a $L^2$ normalization term. 
We define the size penalty as
\begin{equation}\label{eqn:size_penalty}
\xi_{\rm size} = \sigma_{\rm size} + \alpha_{\rm size} \log{e_{\rm size}}    ,
\end{equation}
where $\sigma = \sigma_{k1}^2 + \sigma_{k2}^2$.
In the case of multiple Gaussians, we take the summation.
$e_{\rm size}$ is the excess over an adaptive threshold defined as 
\begin{equation}
    e_{\rm size} = \text{relu}\Big(\sigma_{\rm size} - \max{\big(1.0, \frac{\sigma_{k1} + \sigma_{k2}}{2}}\big) \Big),
\end{equation}
and $\alpha_{\rm size}$ controls how aggressively the threshold is applied. 
The smoothness penalty is calculated using the total variation with cross terms defined as 
\begin{equation}
\begin{aligned}
\xi_{\text{smooth}} =\ & \sum_{i,j} \left| K_{i,j} - K_{i,j+1} \right| \\
&+ \sum_{i,j} \left| K_{i,j} - K_{i+1,j} \right| \\
&+ \frac{1}{2} \left(
    \sum_{i,j} \left| K_{i,j} - K_{i+1,j+1} \right|
  + \sum_{i,j} \left| K_{i+1,j} - K_{i,j+1} \right|
\right),
\end{aligned}
\end{equation}
where the first term represents the horizontal total variation, the second term represents the vertical total variation, and the third term represents the diagonals of the total variation.
The $L^2$ norm is defined over the $dx$ and $dy$ parameters to ensure that, in the case of multiple Gaussian components, the kernel is maximally clustered at the center of the kernel. The complete prescription is as follows:
\begin{equation}
\begin{aligned}
    \xi_{L^2} = 
    & \sum_{j=1}^N(dx_{1,j} - \mu_{dx_{1}})^2 + \sum_{j=1}^N(dy_{1,j} - \mu_{dy_{1}})^2 \\
    & + \sum_{j=1}^N(dx_{2,j} - \mu_{dx_{2}})^2 + \sum_{j=1}^N(dy_{2,j} - \mu_{dy_2})^2,
\end{aligned}
\end{equation}
where $j$ indexes over the Gaussian components, $N$ is the number of Gaussian components, and $\mu_{dx}$ and $\mu_{dy}$ represent the mean of the respective component. As a reminder, the 1 and 2 represent the first and second kernel, $K_1$ and $K_2$.

\subsubsection{Implementation}

To solve Equation \ref{eqn:langevin} and update the parameter solutions, we must calculate the gradients of $\pi(\vec{x})$ with respect to each free parameter; we do this using the second-order accurate central differences method as implemented in the Python module \texttt{PyTorch} (\citealt{paszke_pytorch_2019}). We apply sigma clipping above 1-$\sigma$ to avoid exploding gradients during the calculation.
We use the Adam optimizer with an initial learning rate of $1e^{-2}$, $\beta_1=0.9$, $\beta_2=0.999$, and $\epsilon=1 \times 10^{-7}$ (\citealt{kingma_adam_2015}).
To ensure appropriate exploration of the parameter space, we set $\zeta=0.01$ and $\tau=0.001$ in Equation \ref{eqn:langevin}.  Additionally, we use a cosine annealing scheduler for both the learning rate and the Langevin temperature with a period of 500. 

In order to help drive convergence, we initialize $K_1=\mathrm{PSF}_1$ and $K_2=\mathrm{PSF}_2$ since this is a trivial (albeit non-ideal) solution to Equation \ref{eqn:primary}.
We initially solve for $\mathrm{PSF}_1$ and $\mathrm{PSF}_2$ by using a simple gradient descent algorithm implemented in \texttt{PyTorch} that uses the Adam optimizer.
We also impose bounds on the parameters so that the $q$ values cannot exceed 1, $0\leq\theta\leq360$, and we bound the sigma values by their initial values. In doing so, we ensure that the solution can not result in larger kernels. 
Without this small tweak, high background (or minimal) values in the ePSF would dominate the convolution.

Finally, in order to reduce the number of time steps required to reach convergence, we apply a preconditioning scheme (\citealt{li_preconditioned_2015}). 
We define the preconditioner at the current time step, $G^t$, as
\begin{equation}\label{eqn:preconditioner}
    G^t = \alpha_G G^{t-\tau} + (1-\alpha_G)\cdot{}\nabla\pi(\vec{x})^2
\end{equation}
where $\alpha_G$ is the preconditioner's scaling term, which we set to 0.9. We initialize $G$ as a vector of zeros.
Therefore, the preconditioned version of Equation \ref{eqn:langevin} is
\begin{equation}
    \phi_i^{t+\tau} = \phi_i^t + \frac{\tau}{\sqrt{G}+\epsilon}\nabla\pi + \zeta\sqrt{2\tau TG}.
\end{equation}
$\epsilon$ is a numeric stabilization value set to $1\times10^{-10}$. 
We note that we also tried a momentum-based technique shown in \cite{kim_stochastic_2022}; however, the preconditioner worked best without the momentum. Our implementation can be found at \href{https://github.com/DragonflyTelescope/dfpsf}{\faicon{github}\texttt{DragonflyTelescope/dfpsf}}.

\subsection{Demonstration on a Toy Problem}

We demonstrate the efficacy of our methodology by applying it to several toy problems. 

\subsubsection{Example 1: An elongated PSF}\label{sec:toy1}
In this example, we begin with a PSF, $\mathrm{PSF}_1$, which is a simple circular Gaussian with $\sigma=1$, $q=1$, and $\theta=0$. $\mathrm{PSF}_2$ is created by convolving $\mathrm{PSF}_1$ with an elliptical Gaussian that has $\sigma=1$, $q=0.6$, and $\theta=40$.  
In this example, we set the number of steps to 2000 in order to allow for the solution to converge; this value was chosen experimentally. We note that allowing the algorithm to run for more iterations does not change the solution.
We also set the solution to be parametrized by a single Gaussian component. We note that allowing the solution to contain multiple Gaussian components does not change the solution; in that case, additional component weights are fit to zero. We leave the kernel size to be equal to the original PSF size ($15 \times 15$ pixels). 

The algorithm takes approximately 30 seconds on a single Intel® Core™ i9-14900HX. Comparatively, when run on a GPU, an NVIDIA GeForce RTX 4060, the computation takes 25 seconds. Since the individual computations for a single step in the algorithm are not computationally expensive, the algorithm does not benefit greatly from using a GPU. Moreover, since each step depends on the result of the previous step, we are unable to further parallelize the code beyond vectorized computations in the step loop.

\begin{figure*}
    \centering
    \includegraphics[width=0.98\linewidth]{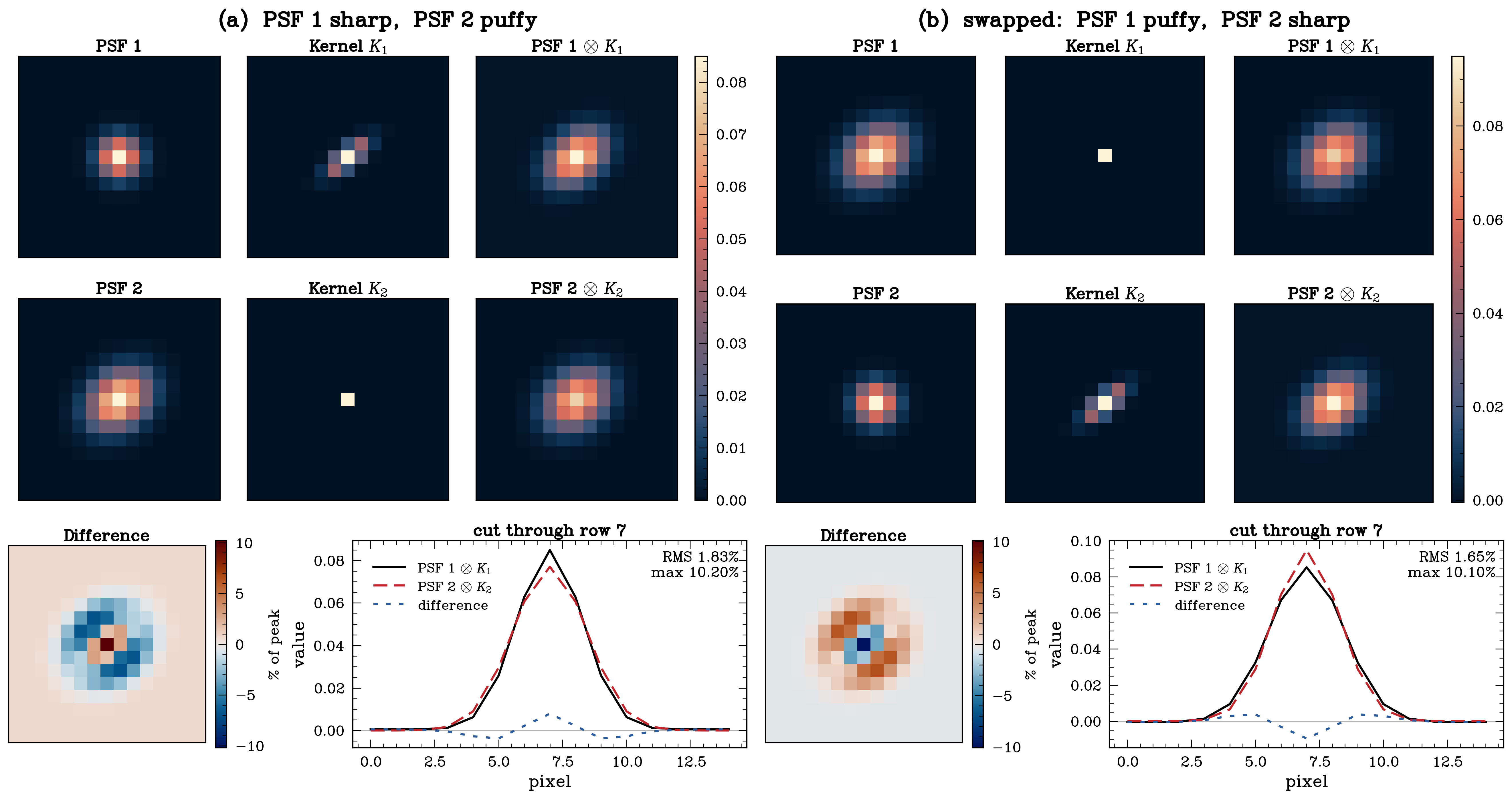}
    \caption{Demonstration of the kernel solutions and final convolutions for $\S$ \ref{sec:toy1}. The left-hand sub figure (a) shows the result if $\mathrm{PSF}_2$ is a convolved version of $\mathrm{PSF}_1$. The right-hand sub figure (b) inverts $\mathrm{PSF}_1$ and $\mathrm{PSF}_2$ in order to demonstrate that the method is invariant to the PSF order. In both cases, the algorithm recovers the original kernel used to convolve the original PSF and finds a delta function for the already convolved PSF (i.e., in the left-hand pair, SGLD correctly calculates kernel 1 to be the original kernel used to create $\mathrm{PSF}_2$ from $\mathrm{PSF}_1$ and kernel 2 to be the delta function).  In each sub figure we show, in the top row, the original $\mathrm{PSF}_1$, the solved-for kernel, $K_1$, and the result of $K_1$ convolved with $\mathrm{PSF}_1$. In the second row we show the original PSF2, the solved-for kernel, $K_2$, and the result of $K_2$ convolved with $\mathrm{PSF}_2$. The bottom row shows the difference between $\mathrm{PSF}_1 \otimes K_1$ and $\mathrm{PSF}_2 \otimes K_2$ and the value of the difference as a function of the pixel taken as a cut through row 7. We also report the RMS of the difference and the max error compared to the peak.}
    \label{fig:toy1}
\end{figure*}

On the left-hand side of Figure \ref{fig:toy1}, we show the results of the DKO framework on this test. In each sub figure, we show, in the top row, the original PSF1, the solved-for kernel, $K_1$, and the result of $K_1$ convolved with PSF1. In the second row we show the original PSF2, the solved-for kernel, $K_2$, and the result of $K_2$ convolved with PSF2. The bottom row shows the difference between $\mathrm{PSF}_1 \otimes K_1$ and $\mathrm{PSF}_2 \otimes K_2$ and the value of the difference as a function of the pixel taken as a cut through row 7. We also report the RMS of the difference and the max error compared to the peak.
We remind the reader that the goal of the DKO framework is to match these convolutions such that the kernels are as compact as possible.
On the right-hand side, we show the same results except we have swapped the roles of $\mathrm{PSF}_1$ and $\mathrm{PSF}_2$ in order to demonstrate that the algorithm is invariant to the order of the PSFs.

The difference images show that residual structure remains even after applying the DKO framework; however, the structure of the convolved PSFs do match qualitatively to around 10\%.

\subsubsection{Example 2: Identical PSFs}\label{sec:toy2}
\begin{figure}
    \centering
    \includegraphics[width=0.95\linewidth]{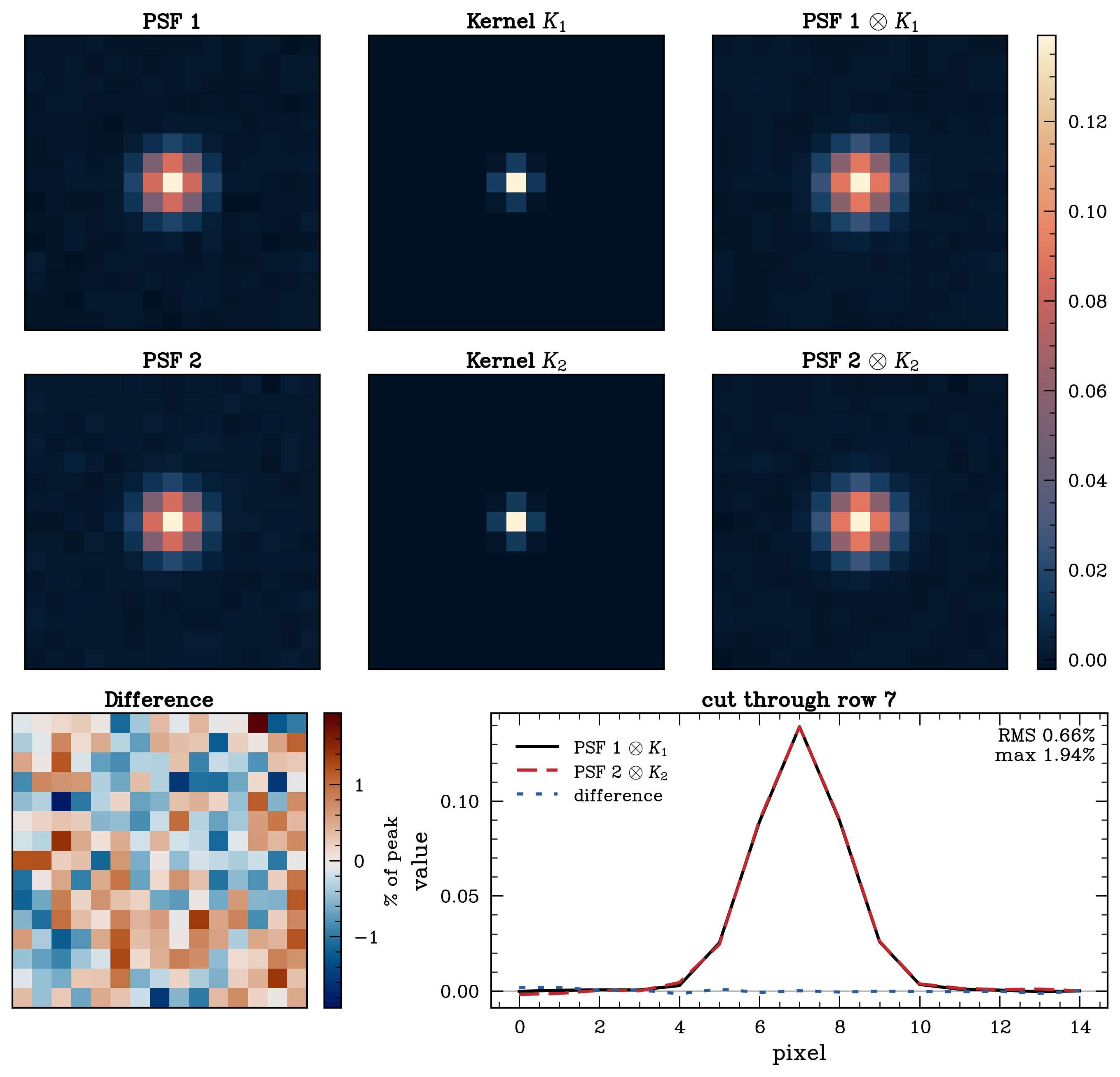}
    \caption{Demonstration of the PSFs and kernel solutions discussed in $\S$ \ref{sec:toy2}. In each sub figure we show, in the top row, the original $\mathrm{PSF}_1$, the solved-for kernel, $K_1$, and the result of $K_1$ convolved with $\mathrm{PSF}_1$. In the second row we show the original $\mathrm{PSF}_2$, the solved-for kernel, $K_2$, and the result of $K_2$ convolved with $\mathrm{PSF}_2$. The bottom row shows the difference between $\mathrm{PSF}_1 \otimes K_1$ and $\mathrm{PSF}_2 \otimes K_2$ and the value of the difference as a function of the pixel taken as a cut through row 7. We also report the RMS of the difference and the max error compared to the peak. Since the PSFs are identical, we expect the solved-for kernels to be identical and equal to unity. Although the kernels deviate slightly from delta functions, the convolved PSFs do not show any notable increase compared with the original PSFs.}
    \label{fig:toy2}
\end{figure}

In this next example, we take two identical PSFs defined as Gaussians with $\sigma=1$, $q=1$, and $\theta=0$. We add a small amount, $\mu=0.01$, of additive Gaussian noise. The goal is to demonstrate that the algorithm recognizes that the optimal solution is a single pixel kernel with a value near unity. In the case with no noise, the single pixel kernel would be exactly at unity. 
In the bottom row of Figure \ref{fig:toy2}, we see that the difference image of the two convolved kernels is noise on the 1\% level. This indicates that the two convolved kernels match exactly to the additive noise level injected into the original kernels. Although this is a highly contrived toy problem, it shows that the DKO framework will collapse towards an optimal solution in the trivial case.

\subsubsection{Example 3: Distinct PSFs with Noise}\label{sec:toy3}

\begin{figure}

	\includegraphics[width=0.95\linewidth]{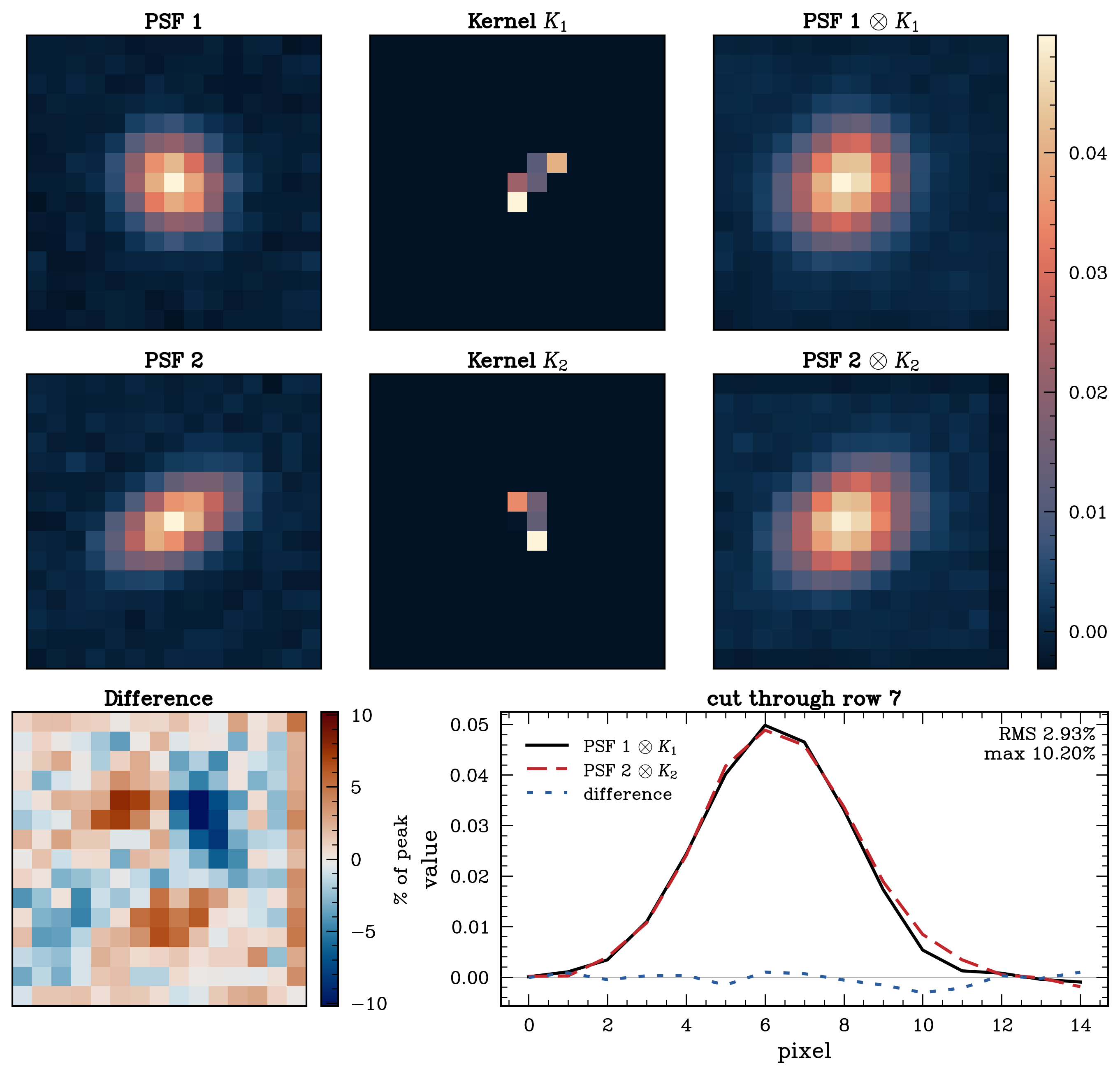}
    \caption{Toy problem demonstrating the efficacy of the SGLD approach in recovering kernels when using a mixture of 2 Gaussians to model each kernel corresponding to $\S$ \ref{sec:toy3}. 
    In each sub figure we show, in the top row, the original $\mathrm{PSF}_1$, the solved-for kernel, $K_1$, and the result of $K_1$ convolved with $\mathrm{PSF}_1$. In the second row we show the original $\mathrm{PSF}_2$, the solved-for kernel, $K_2$, and the result of $K_2$ convolved with $\mathrm{PSF}_2$. The bottom row shows the difference between $\mathrm{PSF}_1 \otimes K_1$ and $\mathrm{PSF}_2 \otimes K_2$ and the value of the difference as a function of the pixel taken as a cut through row 7. We also report the RMS of the difference and the max error compared to the peak.}
    \label{fig:toy_noise4}
\end{figure}

The setup of the problem is as follows: we start with a compact circular Gaussian with $\sigma=1$, $q=1$, and $\theta=0$. To construct $\mathrm{PSF}_1$ and $\mathrm{PSF}_2$, we apply two different convolutional kernels. The first has $\sigma=1.2$, $q=0.8$, and $\theta=-45$. The second has $\sigma=1$, $q=0.3$, and $\theta=30$. Although this situation is not likely to arise naturally in astronomical imaging, this toy problem is meant to challenge the algorithm. We then apply uniform Gaussian noise across the PSF with a mean of 0.02.
In Figure \ref{fig:toy_noise4}, we show the PSFs which will be used in Equation \ref{eqn:primary} (row 1). In row 2, we show our computed solutions, $K_1$ and $K_2$. In the last row, we show the result of $\mathrm{PSF}_1\otimes K_1$ and $\mathrm{PSF}_2\otimes K_2$.
In this example, we set the number of Gaussian components to 2 for each kernel. Although the kernels computed by the DKO framework are unusual in shape, they result in convolutions that qualitatively match in shape. The difference plot shows that while the center of the convolutions match to the noise level, there are still residuals in the wings.


\section{Application to Data}\label{sec:application}

\subsection{Example 1: Dragonfly SLM PSFs}\label{sec:demo-1}
In the following example, we demonstrate the applicability of the algorithm to image cutouts taken using different filters with highly different PSFs. The images were taken using the Dragonfly Spectral Line Mapper \citep{abraham_distributed_2022, chen_dragonfly_2024, abraham_ultra_2014} located at New Mexico Skies in the United States. The images were taken on 11-23-2024 as part of a set of verification observations intended to test a new camera system. These images were chosen for this example since their respective PSFs are notably different. The images were taken using the \oiii{} narrowband filter and the \oiii{} notched continuum filter. We use $400 \times 350$ pixel cutouts for the two frames.

We construct an ePSF (effective PSF) for both images following the procedure outlined in section \ref{sec:psf-algo}.
The ePSF corresponding to the narrowband image is highly elliptical with a near -45 degree phase angle. The ePSF corresponding to the continuum filter is less elongated with a shallower phase angle. Using the ePSFs for both images, we run the DKO framework to compute the optimal convolutional kernels in order to match the ePSFs. We set the kernels to be represented by 2 Gaussian components and set the kernel size to $15 \times 15$ pixels. We train the algorithm for 8,000 steps which takes approximately 60 seconds using a single Intel® Core™ i9-14900HX. 

\begin{table}
    \centering
    \caption{PSF-matching accuracy on the demo pairs.  Lower is better in every column; the best value per demo is bold. RMS emphasizes the PSF core, MAE weights all residuals linearly, and $\Delta$EE measures redistribution of flux.}
    \label{tab:psf_matching}
    \begin{tabular}{lrrr}
        \toprule
        Method & RMS & MAE & $\Delta\mathrm{EE}$ \\
        \midrule
        \multicolumn{4}{l}{\textit{Demo 1: Dragonfly [O\,\textsc{iii}] vs.\ continuum}} \\
        DKO & $\mathbf{2.50 \times 10^{-4}}$ & $\mathbf{1.70 \times 10^{-4}}$ & $\mathbf{9.74 \times 10^{-3}}$ \\
        photutils & $1.83 \times 10^{-3}$ & $5.65 \times 10^{-4}$ & $8.27 \times 10^{-2}$ \\
        PyPHER & $1.13 \times 10^{-3}$ & $4.01 \times 10^{-4}$ & $7.60 \times 10^{-2}$ \\
        LSST delta-function & $8.86 \times 10^{-4}$ & $3.64 \times 10^{-4}$ & $2.72 \times 10^{-2}$ \\
        Cross-convolution & $6.32 \times 10^{-4}$ & $4.82 \times 10^{-4}$ & $1.49 \times 10^{-1}$ \\
        Tikhonov & $8.76 \times 10^{-4}$ & $3.91 \times 10^{-4}$ & $3.66 \times 10^{-2}$ \\
        \midrule
        \multicolumn{4}{l}{\textit{Demo 2: Dragonfly vs. WISE}} \\
        DKO & $\mathbf{2.82 \times 10^{-5}}$ & $\mathbf{1.82 \times 10^{-5}}$ & $2.53 \times 10^{-2}$ \\
        photutils & $4.93 \times 10^{-4}$ & $5.17 \times 10^{-5}$ & $3.72 \times 10^{-2}$ \\
        PyPHER & $2.23 \times 10^{-4}$ & $2.89 \times 10^{-5}$ & $\mathbf{8.61 \times 10^{-3}}$ \\
        LSST delta-function & $2.71 \times 10^{-4}$ & $5.26 \times 10^{-5}$ & $6.92 \times 10^{-2}$ \\
        Cross-convolution & $5.02 \times 10^{-5}$ & $3.39 \times 10^{-5}$ & $6.05 \times 10^{-2}$ \\
        Tikhonov & $6.64 \times 10^{-4}$ & $7.62 \times 10^{-5}$ & $8.46 \times 10^{-2}$ \\
        \midrule
        \multicolumn{4}{l}{\textit{Demo 3: HST vs. SPITZER}} \\
        DKO & $2.11 \times 10^{-5}$ & $9.29 \times 10^{-6}$ & $2.63 \times 10^{-2}$ \\
        photutils & $2.26 \times 10^{-6}$ & $2.81 \times 10^{-7}$ & $1.56 \times 10^{-4}$ \\
        PyPHER & $\mathbf{2.70 \times 10^{-8}}$ & $\mathbf{5.27 \times 10^{-9}}$ & $\mathbf{3.80 \times 10^{-5}}$ \\
        LSST delta-function & $9.01 \times 10^{-5}$ & $1.85 \times 10^{-5}$ & $1.33 \times 10^{-1}$ \\
        Cross-convolution & $4.43 \times 10^{-5}$ & $1.45 \times 10^{-5}$ & $9.91 \times 10^{-2}$ \\
        Tikhonov & $5.29 \times 10^{-8}$ & $8.46 \times 10^{-9}$ & $3.80 \times 10^{-5}$ \\
        \bottomrule
    \end{tabular}
\end{table}

\begin{figure*}
    \centering

    \begin{subfigure}{0.48\textwidth}
         \centering
        \includegraphics[width=\linewidth]{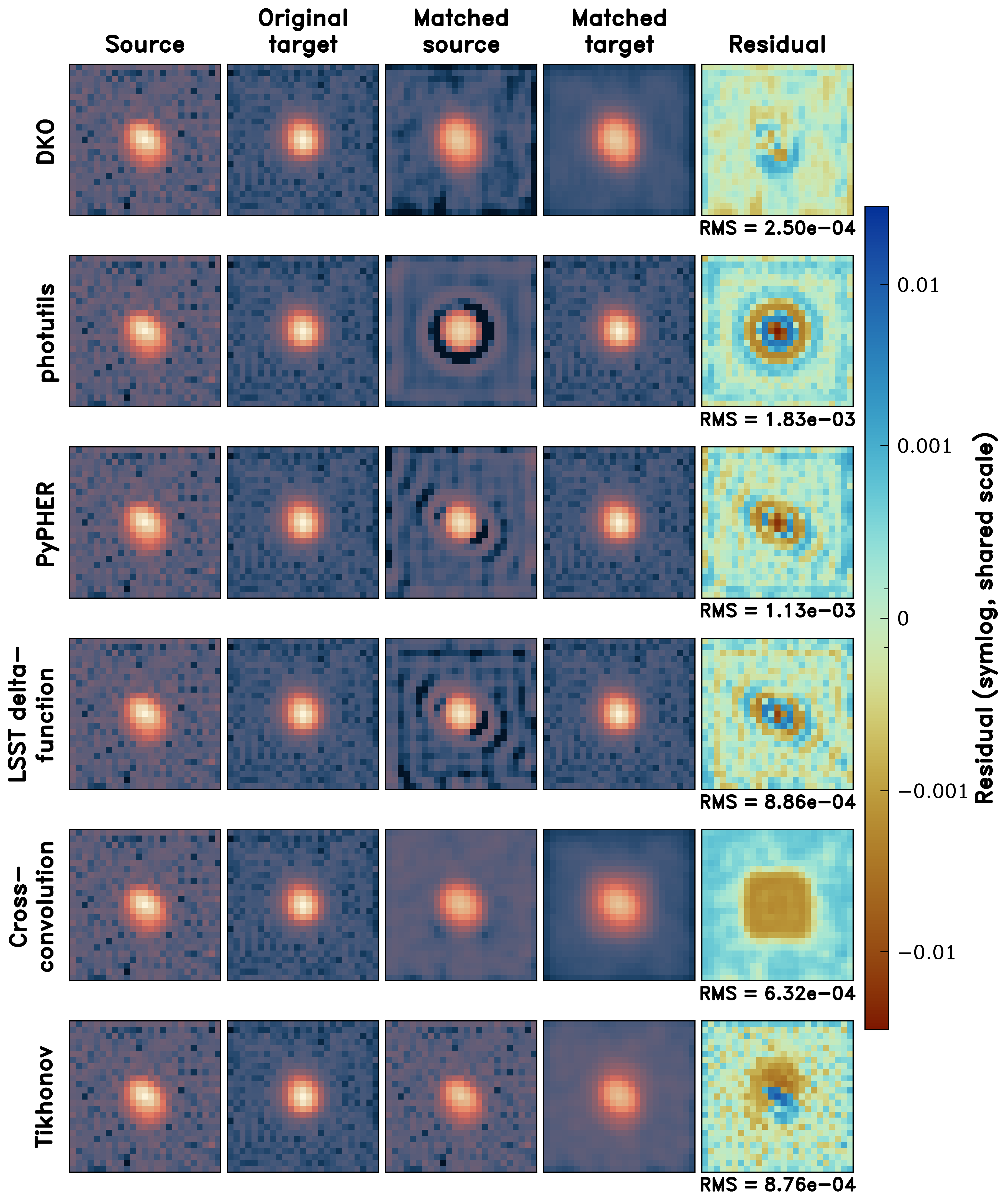}
        \caption{}
    \end{subfigure}
    \hfill
    \begin{subfigure}{0.48\textwidth}
        \centering
        \includegraphics[width=\linewidth]{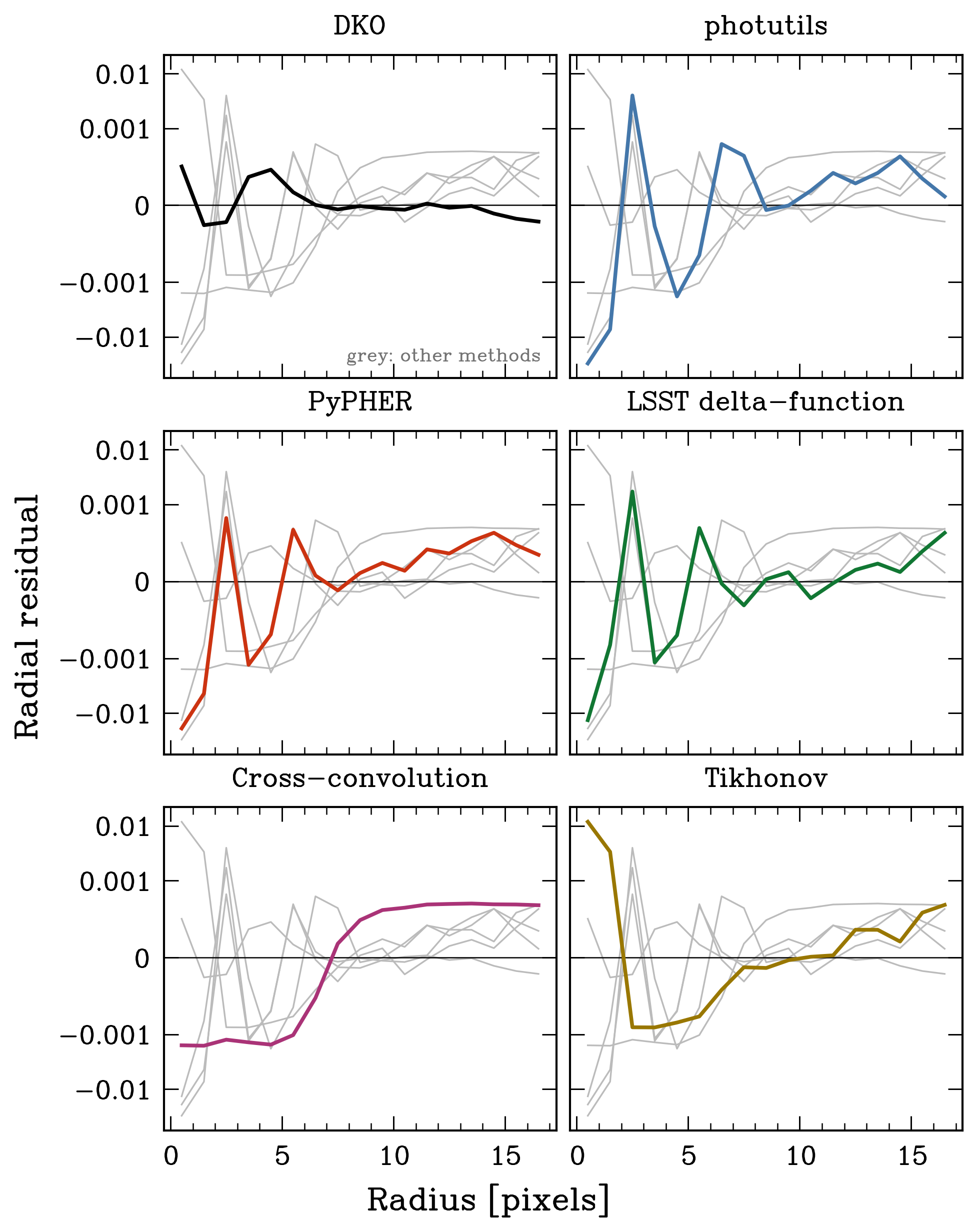}
        \caption{}
    \end{subfigure}

    \caption{\textit{Left}: Source, original target, matched source, matched target, and residual for the 6 methods for comparison on demo 1: DKO, the photutils method, PyPHER, the LSST delta-function method, the cross-convolution method, and the Tikhonov regularization method. 
    The color bar is symlog scaled to bring out both bright and faint 2D structure.
    \textit{Right}: Azimuthally averaged radial residual profiles for the 6 methods. In each panel, the profiles of the other 5 methods are plotted in gray.}
    \label{fig:demo1}
\end{figure*}

In Figure \ref{fig:demo1}(a), we show the original source and target PSFs, the source and target PSFs after convolution, and the residuals for DKO, the photutils method (\citealt{gordon_behavior_2008}; \citealt{aniano_common-resolution_2011}), PyPHER (\citealt{boucaud_convolution_2016}), LSST-style delta function (\citealt{becker_regularization_2012}; \citealt{bramich_new_2008}; \citealt{alard_method_1998}; \citealt{alard_image_2000}), cross correlation (\citealt{yuan_astronomical_2008}), and the Tikhonov regularization method. These methods and their implementations are described in Appendix \ref{app:methods}. While DKO and the cross convolution method solve Equation \ref{eqn:primary}, the other methods solve Equation \ref{eqn:simple-conv}; therefore, it is necessary to select a source and target PSF for these methods. We set the continuum image's PSF as the target PSF since it has a larger FWHM. The figure demonstrates that the three uni-directional PSF matching algorithms, the photutils method, PyPHER, and the LSST delta-function method, all introduce ringing artifacts typically seen as the result of an ill-conditioned matrix inversion. The heavy regularization of the Tikhinov implementation supresses the ringing structure, but the residual shows a strong dipole. The cross-convolution method introduces a strong pedestal while minimizing additional structure in the residual. Comparatively, DKO results in the weakest dipole with no structured noise. Figure \ref{fig:demo1}(b) shows the azimuthally averaged radial residual profiles for each method contrasted with the residuals of the other methods. Similarly, this figure demonstrates that DKO has the least structure in the residuals.

We report the root mean squared error (RMS), mean absolute error (MAE) , and change in enclosed energy ($\Delta$EE) in Table \ref{tab:psf_matching} for all demos. For astronomical PSFs where the majority of the flux is concentrated in the center pixels, the RMS effectively upweights the central pixels since it weights larger residuals more than small residual and thus is a proxy for how well the cores match. On the contrary, mean absolute error is weighted linearly and thus allows the wings of the PSF to contribute more to the metric making it a good proxy for how well the wings match. Finally, the enclosed energy measures the flux redistribution throughout the entire PSF; therefore, $\Delta$EE indicates if any flux distribution is lost through the matching process. Individually, these metrics show an incomplete picture of the ability of an algorithm to match PSFs. Taken together, in addition with the 2D residual structure, they can discriminate between good and bad algorithms for a given PSF matching problem.
For DSLM, DKO ranks best in all three categories. Most importantly, it has the lowest $\Delta$EE value indicating that flux is not redistributed from one region of the PSF to another.

\subsection{Example 2: Dragonfly and WISE}

In this example, we compare the different methods of matching PSFs for a PSF constructed from Dragonfly observations and a WISE W3 PSF. The Dragonfly PSF was constructed from g-band observations presented in \cite{pasha_bullseye_2025} using the PSF construction methodology outlined in \cite{liu_fuzzy_2025}. The WISE W3 PSF is the central focal plane PSF taken from the WISE empirical PSF library. The PSF was constructed following the standard WISE pipeline for single-exposure PSFs.

The Dragonfly PSF is relatively compact (FWHM $\approx$ 2 pixels) with an extended wing spanning several pixels radially. There are faint diffraction spikes oriented vertically and horizontally. By comparison, the WISE W3 PSF has a FWHM $\approx$ 5 pixels and strong diffraction spikes horizontally, vertically, and diagonally. Since the WISE W3 PSF has a larger FWHM, we treat it as the target PSF.
For our algorithm, we set the kernels to be represented by 2 Gaussian components and set the kernel size to $18 \times 18$ pixels; the algorithm is trained for 5,000 steps.

\begin{figure*}
    \centering

    \begin{subfigure}{0.48\textwidth}
         \centering
        \includegraphics[width=\linewidth]{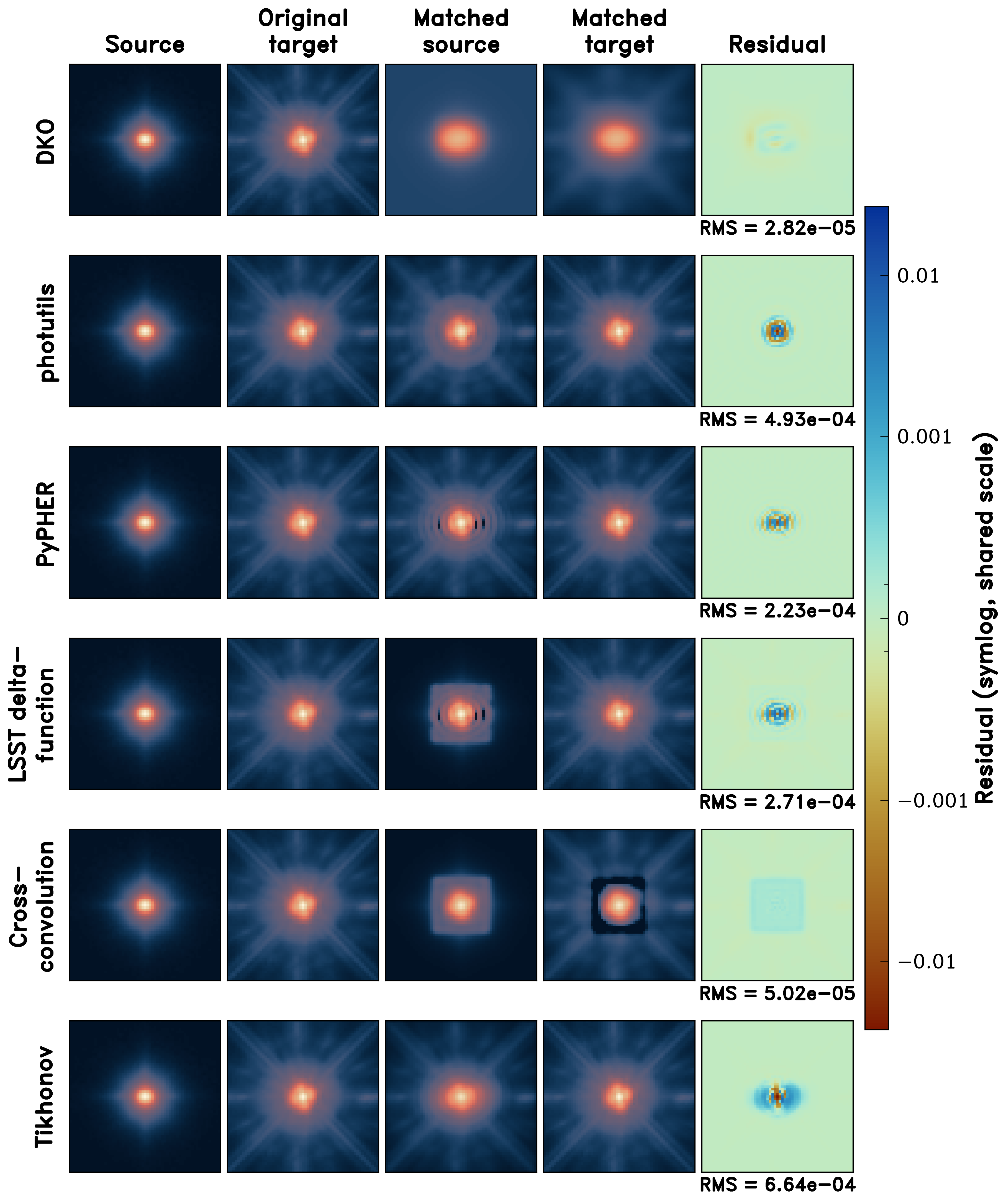}
        \caption{}
    \end{subfigure}
    \hfill
    \begin{subfigure}{0.48\textwidth}
        \centering
        \includegraphics[width=\linewidth]{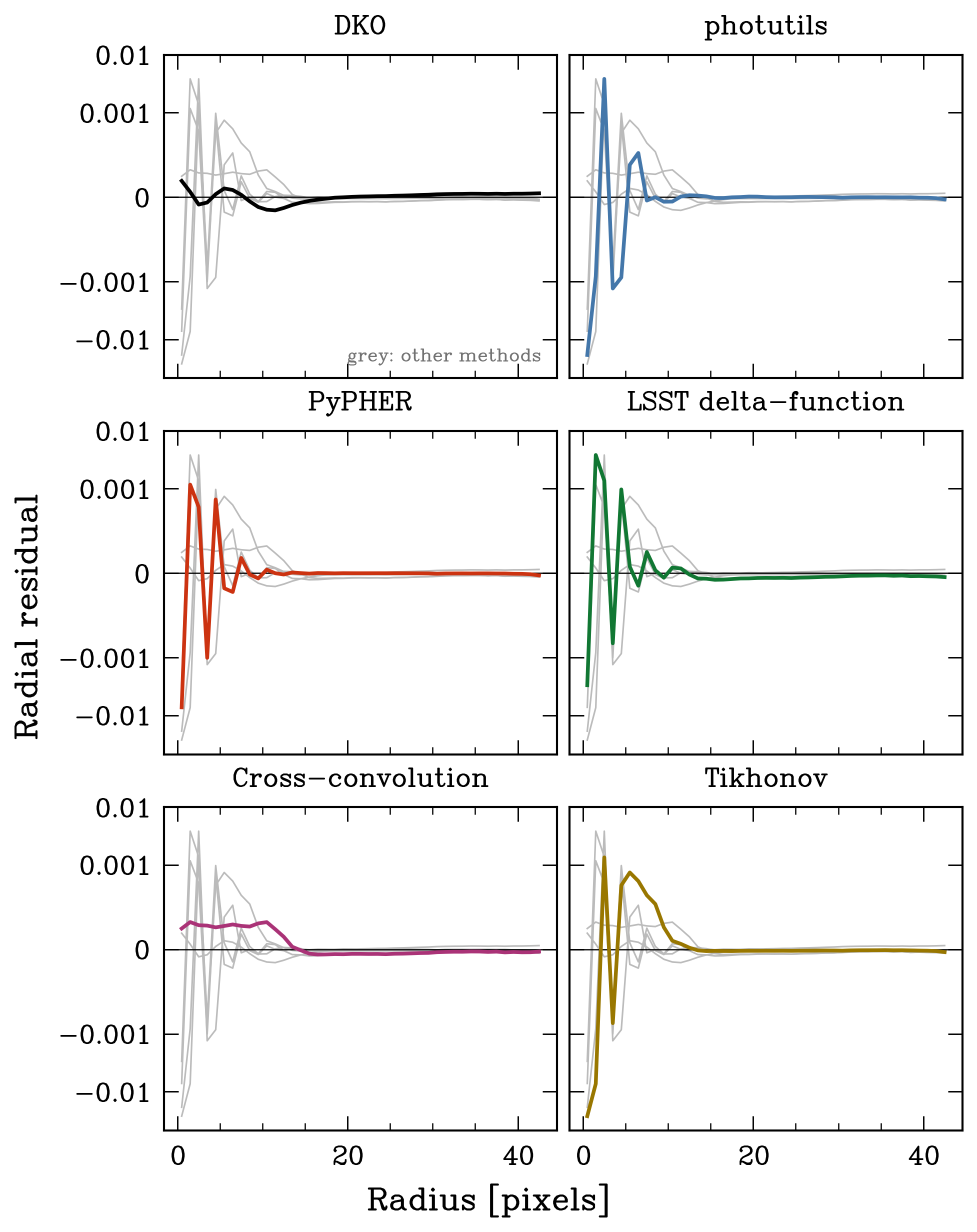}
        \caption{}
    \end{subfigure}

    \caption{Same as Figure \ref{fig:demo1} but for demo 2, Dragonfly and WISE.}
    \label{fig:demo2}
\end{figure*}

In Figure \ref{fig:demo2}a, we show the source, target, matched source, matched target, and residual plots for the 6 methods. The RMS, MAE, and $\Delta$EE values are reported in table \ref{tab:psf_matching}. All four uni-directional methods introduce strong ringed artifacts in the PSF; however, they capture the structure of the diffraction spikes. Although the PyPHER method shows strong ringed artifacts in the residual, it obtains the lowest $\Delta$EE. The cross-correlation algorithm accurately matches the center of the PSFs but introduces a strong pedestal and misses the diffraction spikes. Finally, DKO smears out the diffraction spikes in the matched target and does not add them to the matched source. The matched PSFs' cores are larger than the original source or target PSFs; despite a low-level dipole, the matched PSFs match closely. Figure \ref{fig:demo2}b further demonstrates that DKO leaves a dipole but results in very closely matched PSFs.

\subsection{Example 3: HST and Spitzer}
In this example we compare a PSF from the \textit{Hubble Space Telescope} (\textit{HST}) with one from Spitzer. The \textit{HST} PSF is the empirical STDPSF model for the Advanced Camera for Surveys (ACS) Wide Field Channel (WFC) for the F814W filter \citep{anderson_psfs_2006, anderson_one-pass_2022}, built from post-Servicing Mission 4 observations and distributed as a $101\times101\times90$ cube that samples the PSF at 90 field positions on a $9\times10$ grid across the detector. The model is supersampled $4\times$ relative to the $0.05''$ ACS/WFC pixel, i.e. $0.0125~\mathrm{arcsec}~\mathrm{px}^{-1}$. Because the FWHM varies by only ${\sim}3\%$ across the field, we average over the 90 positions to form a single field-averaged PSF. The Spitzer PSF is the cryogenic high-dynamic-range extended point response function (PRF) for IRAC channel 1 \citep{hora_irac_2012}, released in 2007 and normalized as a calibrated image of Vega; it is a $1281\times1281$ image supersampled $5\times$ relative to the ${\approx}1.22''$ IRAC pixel, giving $0.2447~\mathrm{arcsec}~\mathrm{px}^{-1}$. Note that this product is a PRF rather than a PSF; it already folds in the detector sampling and intrapixel sensitivity variation, which is why its FWHM ($1.88''$) exceeds the value usually quoted for the IRAC channel-1 PSF.
We area-rebin both models onto the IRAC grid, which conserves flux exactly, and crop to a $121\times121$ stamp ($29.6''$) retaining $96\%$ of the IRAC curve of growth. The two PSFs differ in FWHM by a factor of ${\approx}23$ ($0.083''$ against $1.88''$), so on the common grid the HST PSF is unresolved: its FWHM is $0.34$ of a pixel and $80\%$ of its flux lands in the central pixel, making it effectively a delta function. In contrast to the previous two examples there is therefore a clear target PSF: the Spitzer PRF.
 In contrast to the previous two examples, there is therefore a clear target PSF: the Spitzer PSF. The results show that DKO does not perform as well in this regime, while regularized unidirectional methods, specifically PyPHER and Tikhonov, accurately match the two PSFs. Quantitatively, these unidirectional methods achieve substantially lower RMS residuals and encircled-energy differences than DKO. Therefore, when one PSF can naturally be treated as a degraded version of the other, regularized unidirectional convolution methods provide the better solution.

\begin{figure*}
    \centering

    \begin{subfigure}{0.48\textwidth}
         \centering
        \includegraphics[width=\linewidth]{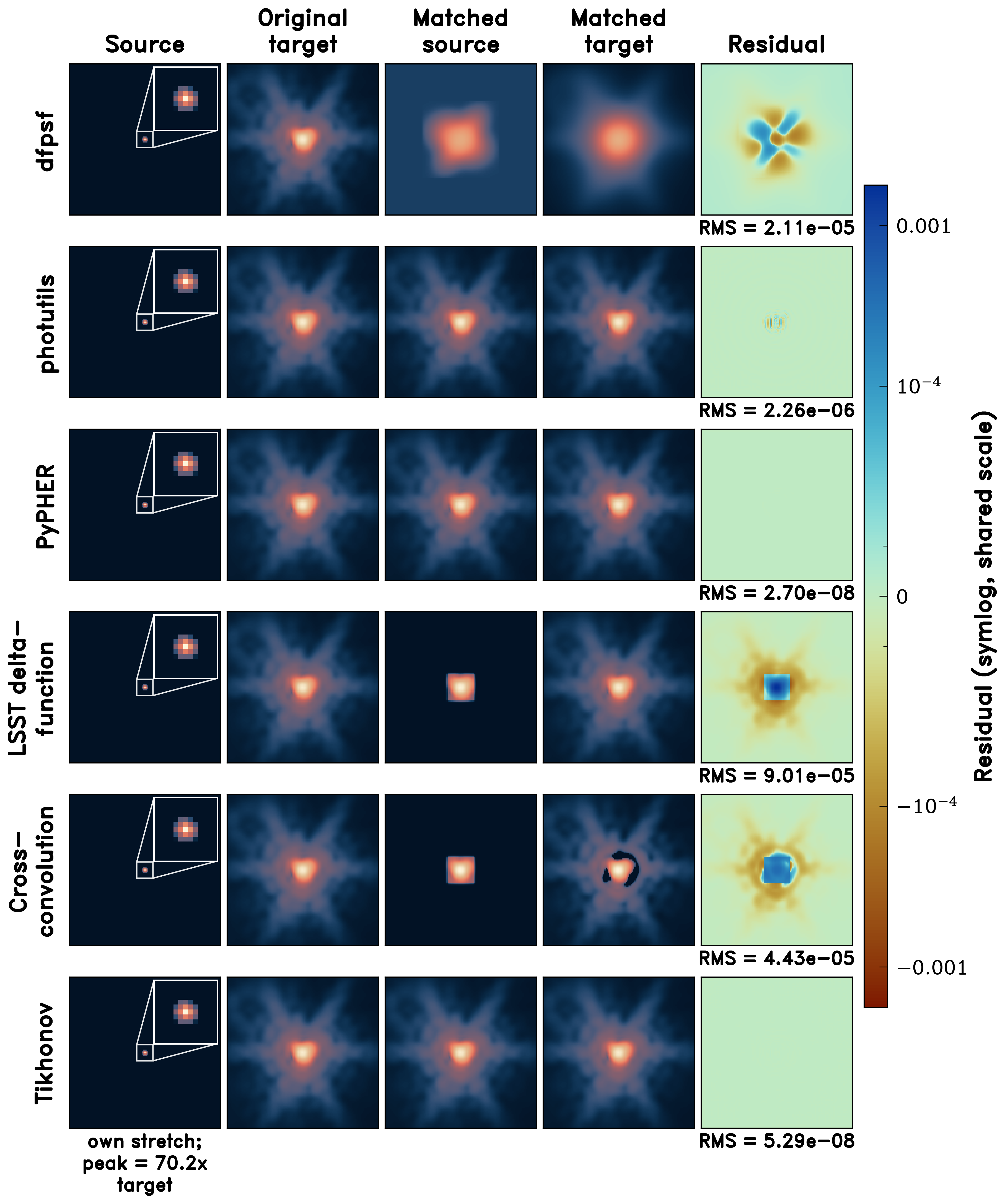}
        \caption{PSFs}
    \end{subfigure}
    \hfill
    \begin{subfigure}{0.48\textwidth}
        \centering
        \includegraphics[width=\linewidth]{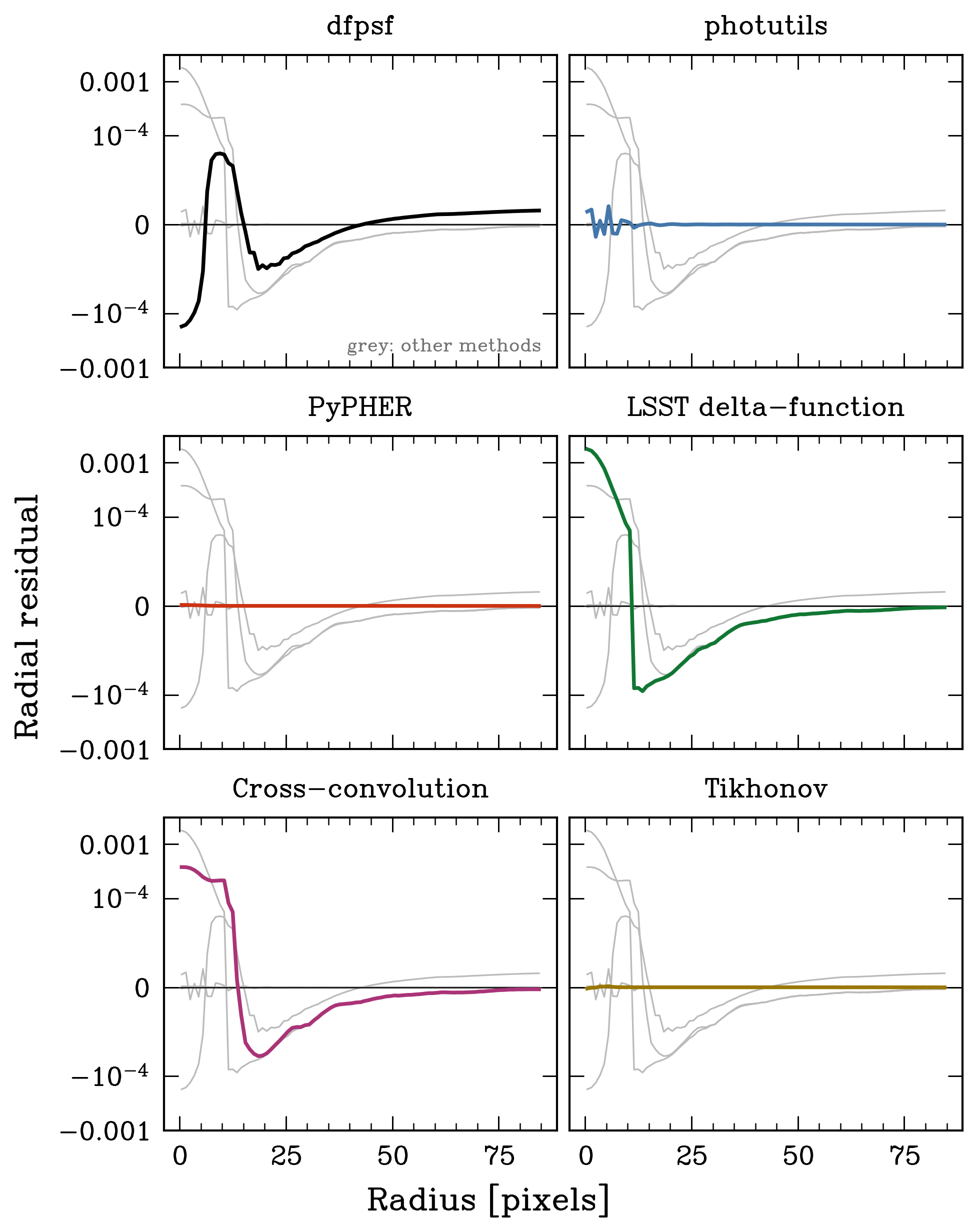}
        \caption{Radial residuals}
    \end{subfigure}

    \caption{Same as Figure \ref{fig:demo1} but for demo 3, \textit{HST} and Spitzer.}
\end{figure*}

\section{Conclusions}\label{sec:conclusions}
The DKO framework offers a methodology for matching drastically different PSFs while conserving the encircled energy. Instead of treating one PSF explicitly as the degraded version of the other, DKO simultaneously computes two distinct convolutional kernels that, when applied to the respective PSF, results in matching PSFs. The DKO framework performs best in cases where neither PSF can be clearly treated as a degraded version of the other, simpler unidirectional methods are better suited to problems in which one PSF can be naturally transformed into the other through convolution.

In the first demonstration, we show that the DKO framework closely matches the source and target PSFs without introducing strong residuals; this is a case in which neither PSF is clearly a degraded version of the other. In Demo 2, we demonstrate that DKO can closely match PSFs that differ substantially in both size and structure, producing the lowest RMS and MAE of the methods considered. However, this comes at the cost of a larger and smoother final PSF, including the suppression of diffraction-spike structure. Finally, in the third case, we show that DKO is not the preferred approach when one PSF can clearly be treated as a degraded version of the other. Instead, our tests indicate that simpler unidirectional convolution algorithms, such as PyPHER or Tikhonov regularization, are better suited to this regime.

The dual-kernel formulation is inherently degenerate since many pairs of kernels will produce final PSFs that match. By initializing the DKO solution as the opposing PSFs, we ensure that the solution is initially, albeit trivially, correct. We then use the SGLD algorithm to explore and exploit the parameter space in order to converge towards compact matched PSFs. We tailor the loss function to favor solutions that not only produce matching PSFs, but also minimize the size of the convolution kernels, thereby limiting the loss of spatial information.

Overall, DKO provides a complementary approach to traditional PSF-matching techniques for cases in which neither PSF provides a natural target for a unidirectional convolution.

\section*{Acknowledgements}

We acknowledge the Dragonfly FRO and particularly thank Lisa Sloan for her project management skills.

We thank the staff at Obstech observatory for their invaluable help with the construction and operation of MOTHRA.

MOTHRA is made possible by the funding and ongoing support from Alex Gerko, Founder and CEO of XTX Markets.

We use the \texttt{cmcrameri} scientific color maps in our figures (\citealt{crameri_scientific_2023}).

Authors declare no conflict of interest

\section*{Data Availability}

All data in presented in this paper can be found at the related GitHub page: \url{https://github.com/DragonflyTelescope/dfpsf}.



\bibliographystyle{rasti}
\bibliography{dfpsf} 



\appendix

\section{Typical Loss Curves}

\begin{figure}
	\includegraphics[width=\columnwidth]{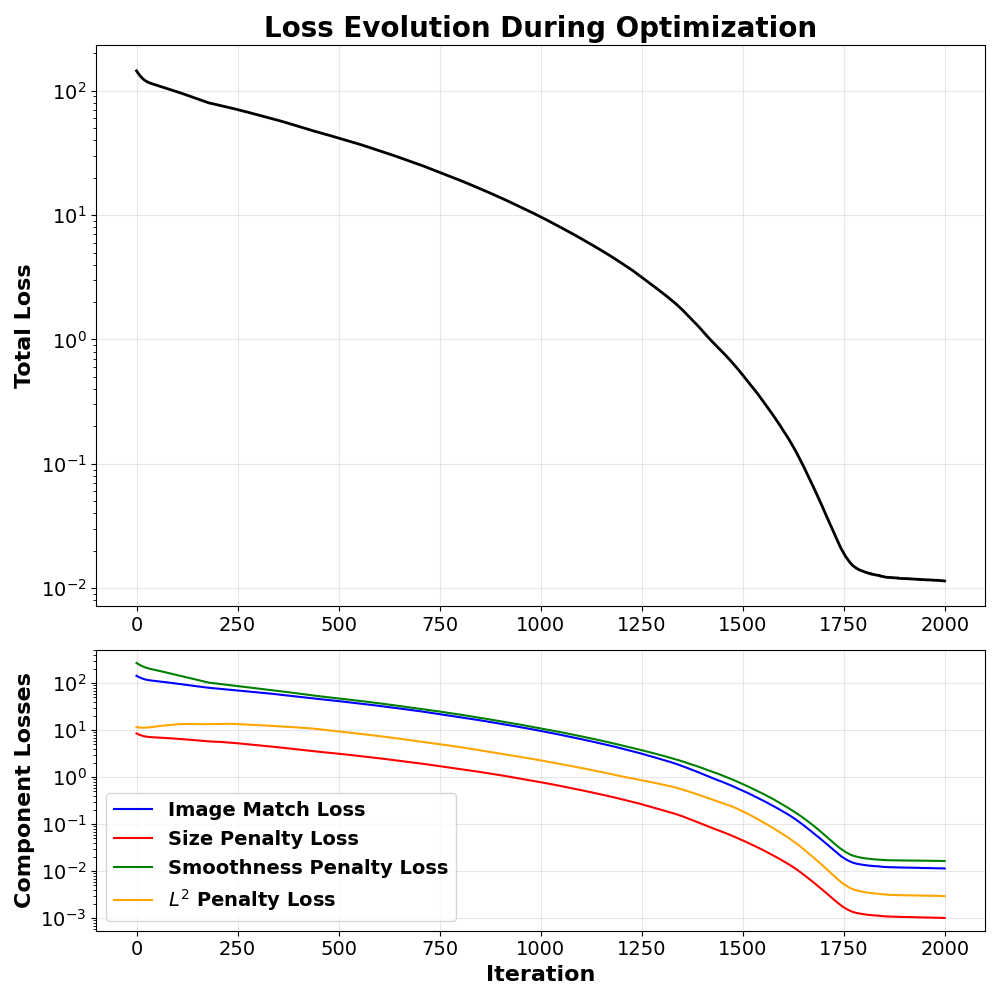}
    \caption{Typical loss curve demonstrating that initially the algorithm is guided by the weighted residual term. The penalty term values here are their dynamically scaled values. In this example after 1700 steps, this term is reduced enough for the size penalty to play the dominate role. Eventually, both terms converge. This behavior is seen consistently in the loss curves.}
    \label{fig:loss_curve}
\end{figure}

Due to how we have structured the loss function, the algorithm first focuses on minimizing the image match loss before minimizing the additional penalty terms. However, since these are all strongly coupled, they all tend to decrease simultaneously. Figure \ref{fig:loss_curve} shows a typical loss curve; this particular loss curve was taken from the example in \S \ref{sec:toy3}.

\section{Constructing an ePSF}\label{app:epsf}

In most applications, the PSF is not immediately known; therefore, we devote the following appendix section to describe our methodology to estimate the PSF which is adopted from photutils. Although we adapt the methodology described in here to the example in subsection \ref{sec:application}, the DKO framework is agnostic to how the ePSFs are created. We emphasize that this methodology is not unique to this paper; we include it for completeness. 
\subsection{Algorithmic Description}\label{sec:psf-algo}
 Therefore, we developed a methodology to build an effective PSF (henceforth the ePSF) given an astronomical image. This process consists of three steps: 
\begin{enumerate}
    \item build a catalog of stars in the image
    \item extract a cutout of each star in the catalog
    \item construct a combined ePSF
\end{enumerate}

To construct the catalog of sources in the image, we use the \texttt{DAOStarFinder} function from the \texttt{Photutils} package \citep{bradley_photutils_2026}. A detailed description of the algorithm can be found in \cite{stetson_daophot_1987}. We set the nominal FWHM to 3.0 and set the roundness to between -0.2 and 0.2. We constrain the sharpness between 0.2 and 1.0. By selecting these bounds, we exclude extended sources from the final catalog. We calculate a background value of the image by taking the sigma-clipped mean. This assumes a constant background over the image; therefore, the images need to be small enough for this assumption to be valid (typically no more than a few hundred pixels by a few hundred pixels). We further remove sources with a flux beneath half the median flux of sources (this removes very faint sources that may have been included). Finally, we discard any blended sources; two sources are considered to be blended if their projected separation is less than 1.5 pixels (half the nominal FWHM). Once curated, we pass the source catalog to the \texttt{extract\_source} function which extracts cutouts around the sources in the catalog from the image. 
The cutouts are supplied to \texttt{EPSFBuilder}. We set the oversampling rate to 1 so that each ePSF will have the same resolution as the images. We apply no smoothing to the ePSF. \texttt{EPSFBuilder} constructs the combined ePSF following the algorithms described in \cite{anderson_toward_2000} and \cite{anderson_empirical_2016}.

\subsection{Demonstration on a Toy Problem}

\begin{figure}
    \centering
    \includegraphics[width=1\columnwidth]{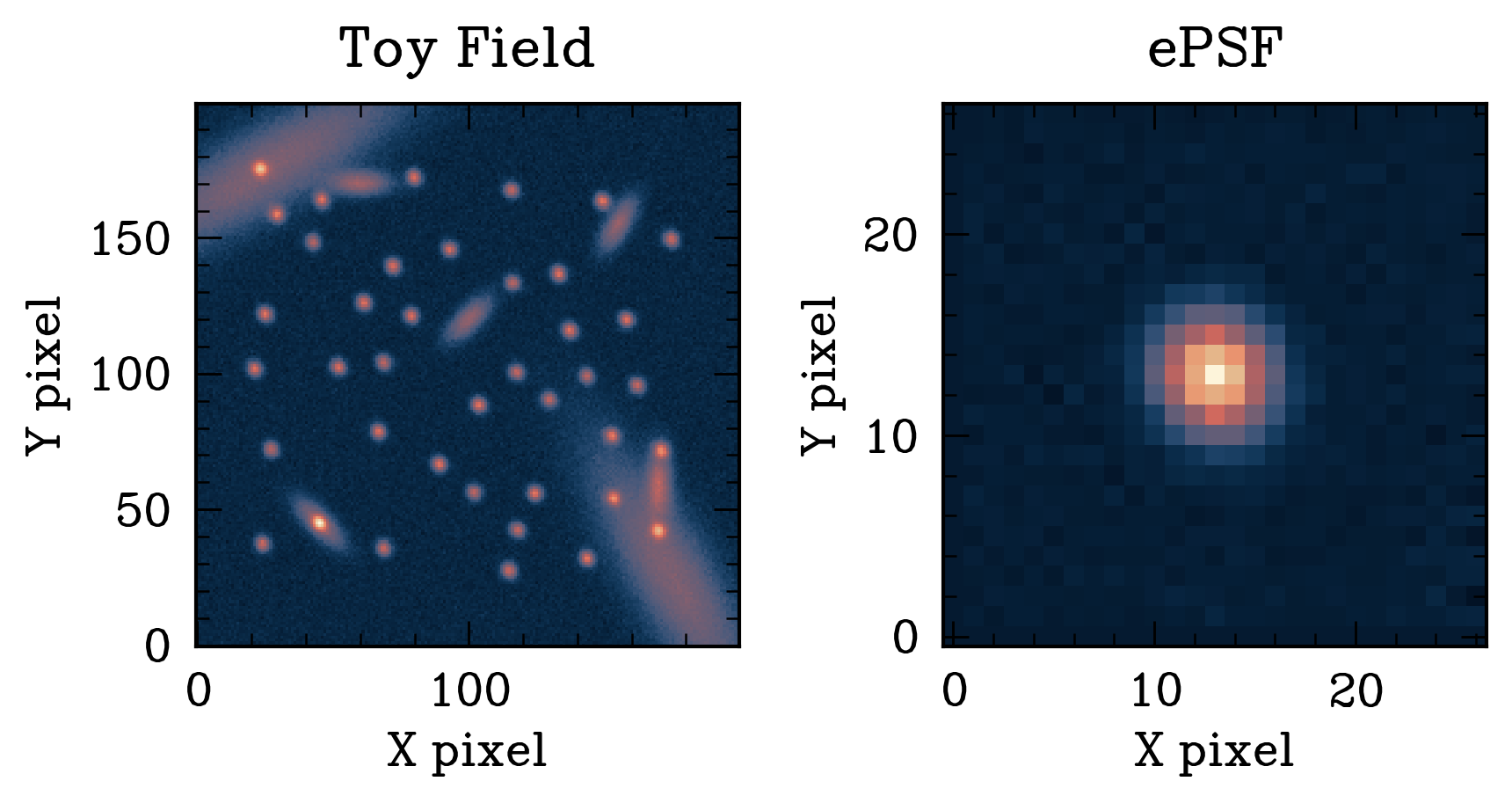}
    \caption{Left: simulated field containing 40 mock stars, 5 mock elliptical galaxies, and 2 mock elongated sources. Right: calculated ePSF using the methodology described in \ref{sec:psf-algo}.}
    \label{fig:toy-paper}
\end{figure}

In order to test this algorithm for building ePSFs, we construct a crowded toy field of $200 \times 200$ pixels containing 40 well-separated Gaussian stars with an elongation of 0.9, a $\sigma$ value of 1.0, a fixed orientation of 30 degrees, and an amplitude between 0.7 and 0.9. In addition to these, we model 5 elliptical galaxies and 2 extended sources. Finally, we add Gaussian noise to the image. We show the toy field and the result of our ePSF creation algorithm in Figure \ref{fig:toy-paper}. Furthermore, we calculate the ellipticity and FWHM of the ePSF. We find an ellipticity of 0.93 and a FWHM of 4.45. These are in close agreement with the values used to simulate the image\footnote{The FWHM of the mock stars are $2.335\sigma = 2.335 * 2 \approx 4.7$.}.

\section{Mathematical Descriptions of Other Solvers}\label{app:methods}
In this appendix, we describe the other methodologies used to solve for matching PSFs. We note that these algorithms were not conceived of by these authors with the exception of the Tikhonov regularization algorithm.

The photutils method, PyPHER, LSST-style delta-function, and Tikhonov regularization methods all solve the 1-dimensional counterpart to equation \ref{eqn:primary}:

\begin{equation}
    \mathrm{PSF}_S \otimes K_1 = \mathrm{PSF}_T
\end{equation}

\subsection{Photutils Method}
This method is based off the the official photutils documentation\footnote{\url{https://photutils.readthedocs.io/en/2.3.0/user_guide/psf_matching.html}}. This is a single-directional PSF matching algorithm in which one PSF is assumed to be the source and the other the target (\citealt{aniano_common-resolution_2011}; \citealt{gordon_behavior_2008}). This method applies the ratio of the Fourier Transforms of the source and target PSFs to the source PSF in order to match it to the target PSF. 

The convolutional kernel is written as 

\begin{equation}
    K(x,y) = \text{FFT}^{-1} \Bigg(
    \frac{\text{FFT}[\text{PSF}_T(x,y)]}{\text{FFT}[\text{PSF}_S(x,y)]} \Bigg)
\end{equation}

However, high frequencies in the Fourier Transform can inject high levels of artificial noise in the result. Therefore, following the recommendation in the documentation, we use a split cosine bell window function with $\alpha=0.3$ and $\beta=0.4$to remove the high frequency noise. 

\begin{equation}
W(\rho) =
\begin{cases}
1, & \rho < r_{\mathrm{in}}, \\[8pt]
\dfrac{1}{2}\left[\,1 + \cos\!\left(\dfrac{\pi\,(\rho - r_{\mathrm{in}})}{\Delta}\right)\right], &
  r_{\mathrm{in}} \le \rho \le r_{\mathrm{in}} + \Delta, \\[8pt]
0, & \rho > r_{\mathrm{in}} + \Delta,
\end{cases}
\label{eq:splitcosbell}
\end{equation}

where $\rho$ is the radial distance to the center of the PSF in pixels, $N=(min(n_x,n_y)-1)/2$, $r_{in}=\beta N$, $\Delta=\lfloor \alpha N \rfloor$. 

The windowed kernel takes the following form:

\begin{equation}
    K(x,y) = \text{FFT}^{-1} \Bigg(
    \frac{\text{FFT}[\text{PSF}_T(x,y)]}{W(\rho)\text{FFT}[\text{PSF}_S(x,y)]} \Bigg)
\end{equation}

\subsection{PyPHER Method}
The PyPHER method solves the uni-direcitonal PSF matching problem by using Wiener filtering with regularization (\citealt{boucaud_convolution_2016}).
The solution to the uni-directional problem is as follows:
\begin{equation}
\widehat{K}(u,v) =
\frac{\widehat{\mathrm{PSF}}_{T}(u,v)\,\widehat{\mathrm{PSF}}_{S}^{\,*}(u,v)}
     {\bigl|\widehat{\mathrm{PSF}}_{S}(u,v)\bigr|^{2} + \mu\,\bigl|\widehat{L}(u,v)\bigr|^{2}} ,
\qquad
L = \begin{pmatrix} 0 & -1 & 0 \\ -1 & 4 & -1 \\ 0 & -1 & 0 \end{pmatrix},
\label{eq:PyPHER}
\end{equation}
where  $\mu$ is the regularization parameter. In our demos, we set this to $1\times10^{-3}$
We compute the convoutional kernel as the real component of the inverse fast Fourier Transform: 
$K = Re[\text{FFT}^{-1}(\hat{K})]$

\subsection{LSST-style Delta Function Method}
The LSST-style delta function method follows the algorithm described in \cite{becker_regularization_2012} which extends the original implementation from \cite{bramich_new_2008}, \cite{alard_method_1998}, and \cite{alard_image_2000}.

The convolutional kernel is written as follows:

\begin{equation}
\hat{\mathbf{K}} = \arg\min_{\mathbf{k}}
  \left\lVert \mathbf{M}\mathbf{k} - \text{PSF}_{T} \right\rVert^{2}
  + \lambda \left\lVert \mathbf{H}\mathbf{k} \right\rVert^{2},
\label{eq:lsst-obj}
\end{equation}

\begin{equation}
\left( \mathbf{M}^{\!\top}\mathbf{M} + \lambda\,\mathbf{H}^{\!\top}\mathbf{H} \right)
  \hat{\mathbf{k}} = \mathbf{M}^{\!\top}\,\mathrm{PSF}_{T},
\label{eq:lsst-normal}
\end{equation}

where $\mathbf{k}$ is the vectorized convolution kernel and
$\mathbf{M}$ is the convolution matrix constructed from the source
PSF, such that $\mathbf{M}\mathbf{k}$ represents
$\mathrm{PSF}_S \otimes K$. The first term minimizes the squared
residual between the convolved source PSF and the target PSF,
$\mathrm{PSF}_T$. The matrix $\mathbf{H}$ is the regularization
operator, which penalizes undesirable structure in the recovered
kernel, and $\lambda$ controls the strength of this regularization.
Equation~\ref{eq:lsst-normal} is the corresponding normal equation
used to solve for the optimal kernel coefficients $\hat{\mathbf{k}}$.

Although, \cite{bramich_new_2008} suggests $0.1<\lambda<1.0$, we use $\lambda=1\times10^{-4}$. We selected this value experimentally since it gave the lowest RMSE on the matched convolutions. 

\subsection{Cross Convolution Method}
The cross convolution method follows the algorithm described in \cite{yuan_astronomical_2008}. The authors use a linear least-squares optimization algorithm which simultaneously solves for the convolutional kernels by minimizing the pixel-level differences and including a regularization term penalizing the radial terms of the kernels. The loss is written as follows:
\begin{equation}
\begin{split}
Q ={}& \sum_{x,y}
\left[(\mathrm{PSF}_1\otimes K_1)(x,y)
-(\mathrm{PSF}_2\otimes K_2)(x,y)\right]^2 \\
&+ \lambda \sum_{x,y}(x^2+y^2)^2
\left[K_1(x,y)^2+K_2(x,y)^2\right].
\end{split}
\end{equation}
While similar to our approach, this method uses a different minimization algorithm and different regularization terms.

\bsp	
\label{lastpage}
\end{document}